\documentclass[]{spie}  

\usepackage{amsmath,amsfonts,amssymb}
\usepackage{graphicx}
\usepackage[colorlinks=true, allcolors=blue]{hyperref}
\usepackage{comment}
\usepackage{wrapfig}
\usepackage{threeparttable}
\usepackage{floatrow}
\usepackage{verbatim}

\newcommand{\aprx}{\mbox{$\ensuremath{\sim}$}}

\title{LuSEE-Night power requirements and power generation strategy}

\author[a]{Benjamin R.~B. Saliwanchik}
\author[a]{Sven Herrmann}
\author[a]{Ivan Kotov}
\author[a]{Paul O'Connor}
\author[b]{Maxim Potekhin}
\author[b]{Anže Slosar}
\author[c,d]{Stuart Bale}

\affil[a]{Instrumentation Division, Brookhaven National Laboratory, Upton, NY, USA}
\affil[b]{Physics Department, Brookhaven National Laboratory, Upton, NY, USA}
\affil[c]{Physics Department, University of California, Berkeley, CA, USA}
\affil[d]{Space Sciences Laboratory, University of California, Berkeley, CA, USA}

\authorinfo{Further author information: (Send correspondence to B.R.B.S.)\\ E-mail: bsaliwanc@bnl.gov}

\begin{document} 
\maketitle

\begin{abstract}

The Lunar Surface Electromagnetics Experiment at Night (LuSEE-Night) is a project designed to investigate the feasibility of observing the Cosmic Dark Ages using an instrument on the lunar far-side. LuSEE-Night will measure the redshifted 21\,cm transition of neutral hydrogen over a frequency range of 0.1-50 MHz, covering the redshift range $27 < z < 1100$. The LuSEE-Night instrument is a radio frequency spectrometer, consisting of four horizontal monopole antennas, arranged to give wide zenith-pointing beams with two orthogonal linear polarizations. This combination of polarization, spectral, and angular sensitivity will be necessary to separate the cosmological signal from significantly stronger foreground emissions. LuSEE-Night will observe in drift scan during lunar night while the moon shields it from radio frequency interference from both the Earth and sun, and will transmit science and telemetry data back to Earth via an orbital relay during the lunar day. LuSEE-Night will have to operate in a challenging environment: its electronics must operate under hard radiation, the instrument must be thermally isolated during the cold 100~K lunar night, and have a thermal rejection path to survive the 390~K daytime temperature, and its photovoltaic and battery systems must provide sufficient power to operate during two weeks of lunar night. Furthermore, the instrument spectrometer must be powered throughout the lunar night using only a 7~kWh battery, due to mass limitations. Here we describe the power generation, storage, and delivery subsystems of the LuSEE-Night instrument, and the simulations of expected power generation, draw, and reserves over time which were performed in order to design the power subsystems, and ensure instrument survival and operation throughout the long lunar night. We also describe the Concept of Operations (ConOps) developed for the LuSEE-Night mission, which derives from the power management simulations.

\end{abstract}

\section{Introduction}

One of the remaining unexplored periods in the universe's history is the Cosmic Dark Ages, the epoch after the universe became transparent to electromagnetic radiation at the surface of last scattering of the Cosmic Microwave Background (CMB), and before the first luminous astrophysical sources formed. The CMB is well constrained to have occurred at \aprx370,000~yrs after the Big Bang, but the period of ``First Dawn'', when the first stars form is less well constrained, but believed to be \aprx200-500~Myrs after the Big Bang. During the intervening Dark Ages, the only baryonic matter in the universe was cold, dark neutral hydrogen gas, which is observable with traditional optical astronomy techniques. Potentially the only mechanism for observing this period is the hyperfine or ``spin-flip'' transition of neutral hydrogen, which produces an absorption/emission line with a frequency of 1.42~GHz, or 21\,cm in wavelength. 

This transition line in principle allows the neutral hydrogen from this epoch to be observed as an absorption/emission feature against the back-light of the CMB. 
The 21\,cm line is a classically forbidden transition, which produces a very fine absorption/emission line, allowing observations to simultaneously measure the redshift to the gas that sourced the line, and therefore build up three-dimensional tomographic maps of the large scale structure at the time of interaction \cite{Chang:2007xk}. In recent years, numerous radio telescopes have been developed to use the 21\,cm line to explore different epochs of the universe's history, including the HERA\cite{deboer17}, CHIME\cite{chime22}, and HIRAX\cite{crichton22} instruments, which focus on the later reionization or Dark Energy dominated epo of the universe, the US Department of Energy's (DOE) Baryon Mapping Experiment (BMX)\cite{oconnor20}, which observes the 21\,cm line in the local universe, and the DOE's proposed PUMA array\cite{bandura19}.

The global, sky-averaged, spectrum of the redshifted 21\,cm line is also sensitive to the temperature and polarization state of the neutral hydrogen in the universe as a function of redshift. Instruments such as EDGES\cite{Monsalve19}, PRIZM\cite{philip19}, and SARAS\cite{singh21} are designed to observe this 21\,cm monopole signal. The monopole signal is significantly easier to detect than a resolved 21\,cm signal. However, observing even the monopole 21\,cm signal from the Cosmic Dark Ages from Earth presents new difficulties.

Due to cosmic expansion, the 21\,cm signal from the Dark Ages is redshifted to the frequency range of approximately 3-30MHz at the present time. This frequency range is difficult to measure from the surface of the Earth for several reasons, including ionospheric opacity and anthropogenic and meteorological sources of radio frequency interference (RFI). The lunar far-side is potentially the best location in the solar system from which to observe in this frequency band, due to the lack of ionosphere, and the bulk of the moon, which can simultaneously shield the observatory from RFI from the Earth and the Sun.

For these reasons, LuSEE-Night was designed as a pathfinder experiment to demonstrate the feasibility of observing the Cosmic Dark Ages from the lunar far-side. LuSEE-Night will benefit from a pristine RFI quiet environment, an expected $>100$~dB of shielding from solar and terrestrial radio emissions in band, and observations without interfering ionospheric distortion or opacity. LuSEE-Night will observe over the frequency range 0.1-50~MHz, which corresponds to a redshift range of $27 < z < 1100$. The goals of the LuSEE-Night instrument are: 1) to establish the feasibility of low frequency radio observations from the lunar far-side, 2) To perform the most sensitive observations of the radio sky in the frequency range 0-50~MHz, 3) to quantify systematic effects on global spectrum measurements, and investigate mitigation methods, and 4) to constrain the presence of a non-smooth monopole signal at the $10^{-3}$ level compared to foregrounds. 

\section{LuSEE-Night Hardware Overview}

\begin{figure}[H]
\begin{center}
\includegraphics[width=0.8 \textwidth]{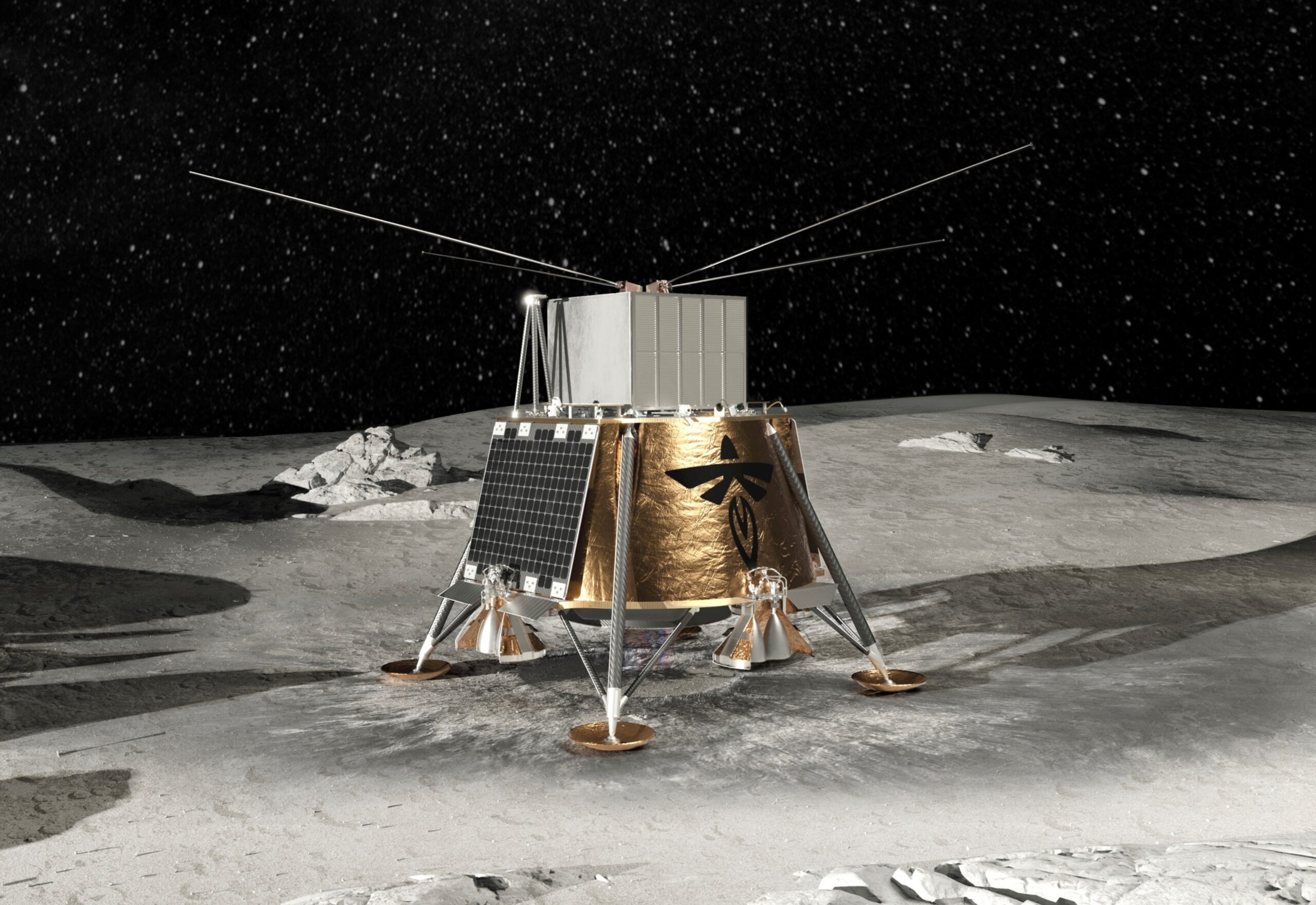}
\end{center}
\caption[]{The LuSEE-Night instrument mounted atop the Blue Ghost Lunar Lander, provided by Firefly Aerospace. The LuSEE-Night instrument consists of an enclosure approximately 1m $\times$ 1m $\times$ 0.7m, and four monopole antennas 3m in length, which operate as two linearly polarized pseudo-dipole antennas. The instrument has a total mass of 128~kg, of which 50~kg is the battery used to power it throughout the 328 hours of lunar night. The battery is charged, and the instrument powered during the lunar day, by three photovoltaic panels, located on the top, east, and west faces of the enclosure. Figure courtesy of Firefly Aerospace.}
\label{fig:lusee_and_lander} 
\end{figure}

The LuSEE-Night instrument will be mounted atop a Blue Ghost Lunar Lander (see Figure \ref{fig:lusee_and_lander}), and delivered to the moon aboard an Elytra Transfer Vehicle, both provided by Firefly Aerospace.\footnote{https://fireflyspace.com/missions/blue-ghost-mission-2/} The Blue Ghost Lunar Lander will land autonomously on the lunar far-side. The lander and all other scientific payloads aboard will permanently shut down at the end of the first solar day after landing on the moon to prevent self-generated RFI from interfering with the sensitive radio observations of LuSEE-Night. All onboard electronics in the LuSEE-Night instrument will be shielded, and clocked to a picket-fence power supply (PFPS) to prevent self-generated RFI.

\begin{figure}[H]
\begin{center}
\includegraphics[width=1.0 \textwidth]{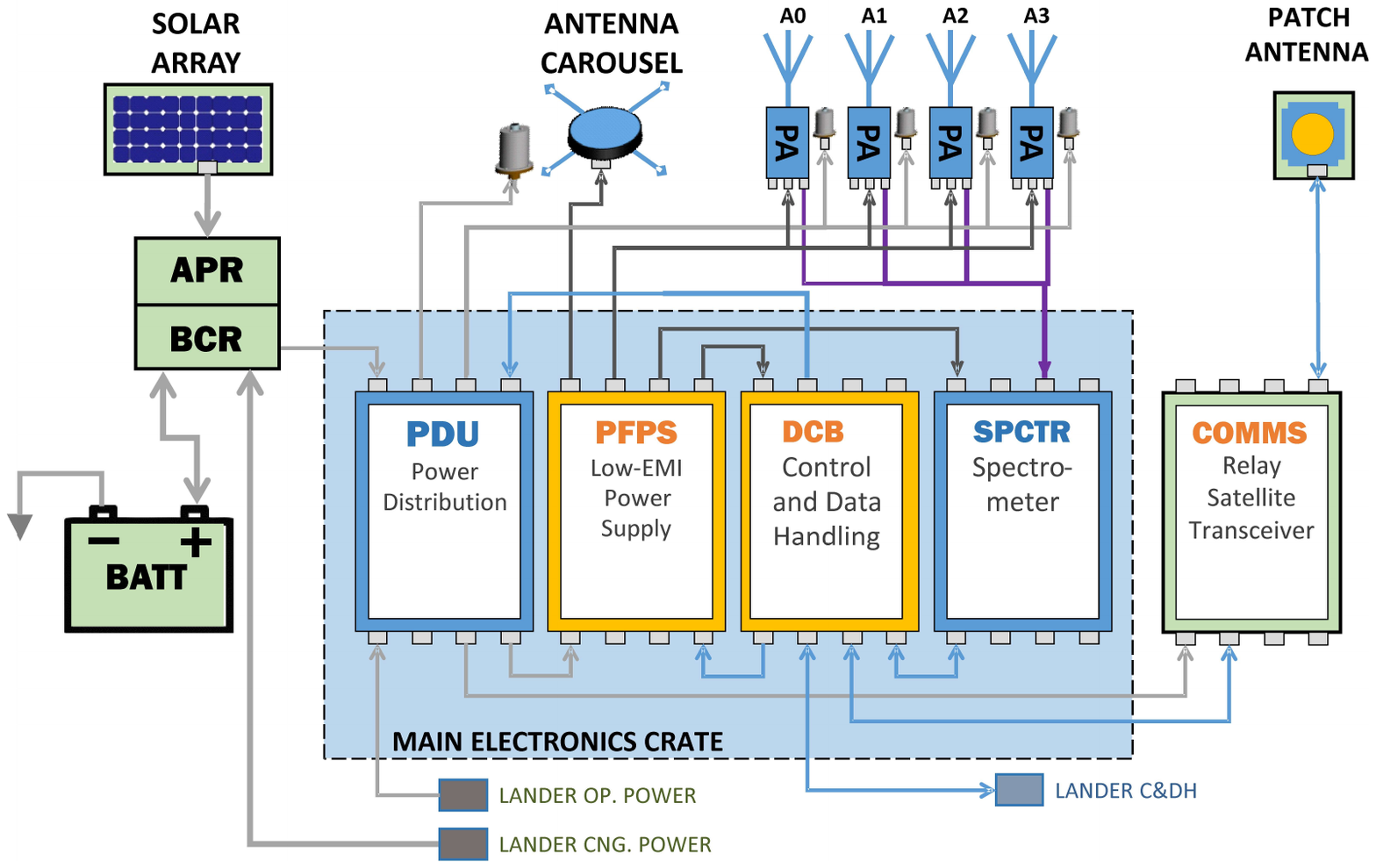}
\end{center}
\caption[]{Block diagram of the LuSEE-Night instrument components and interconnections. Lines in light grey are unregulated power, dark grey are regulated low voltage power, blue are data, and magenta are RF. Components in green are procured from commercial vendors, blue are developed by BNL, and yellow by UC Berkeley Space Sciences Laboratory.
}
\label{fig:lusee_block_diagram} 
\end{figure}

The LuSEE-Night instrument consists of five main components, which are shown in Figure \ref{fig:lusee_block_diagram}, and described below:

\begin{itemize}
    \item The Main Electronics Crate (MEC)
    \item A lithium ion battery, with a nominal capacity of 7160 W-hr
    \item Four 3\,m length stacer monopole antennas, on an azimuthal rotation platform
    \item A photovoltaic solar array to power the instrument during the day and charge the battery for night
    \item An S-band patch antenna for communicating with the communications relay satellite
\end{itemize}

The MEC itself consists of four individually shielded PCB ``slices'': the flight computer or data control board (DCB), the spectrometer board, the power delivery boards, and the communications User Terminal and S-band transceiver. 
The LuSEE-Night spectrometer is described in greater detail in Tamura et al. \cite{tamura24}. The MEC and the lithium ion battery are contained in the Inner Equipment Assembly (IEA), which provides thermal insulation during the lunar night, and heat rejection by a multicell parabolic radiator panel during the lunar day. 

The science and communication antennas and photovoltaic array are located external to the IEA. The monopole science antennas are arranged to give wide zenith-pointing beams with two orthogonal linear polarizations, and are mounted on an azimuthal rotation stage for systematics rejection. The combination of polarization, spectral, and angular sensitivity which the antenna systems will provide are crucial for separating the cosmological signal from the significantly stronger foreground emissions. The full instrument and science are described in Bale et al. \cite{bale23}.

LuSEE-Night will observe the sky in drift scan, and will transmit science and housekeeping data back to Earth via the Lunar Pathfinder (LPF) communications relay satellite.

The entire instrument is limited to a total mass of 128~kg, of which the 7160~W-hr battery accounts for 50~kg. The photovoltaic (PV) array powers the instrument during the lunar day, and charges the battery for nighttime operations. The battery must power the instrument throughout the 328 hours of lunar night, and additionally must keep the battery itself warm enough that it can operate and be recharged when external power from the PV array becomes available again at lunar dawn. Lunar surface temperatures reach as high as 390~K during the day at the LuSEE-Night landing site, while at night they fall as low as 100~K, and the battery temperature must be maintained between approximately $-5^{\circ}$~C and $30^{\circ}$~C (268~K and 303~K). Together, the PV array and battery are limiting factors for the total operating power, and observational duty cycle of the LuSEE-Night instrument, and the simulations of their design and of the instrument operation are the topic of this work.

\section{Solar Array Simulations}
\label{sec:solar_sims}

During operations the power source for LuSEE-Night is the PV array.
The electrical power harvested by the array is stored in the Li-ion battery and distributed to the LuSEE-Night instrument through a Power Distribution Unit (PDU).
To design the layout of the solar array, we conducted extensive simulations of the power delivery subsystem, including numerous properties of the physical system and the operating environment, as described below.

\begin{figure}[H]
\begin{center}
\includegraphics[width=0.49 \textwidth]{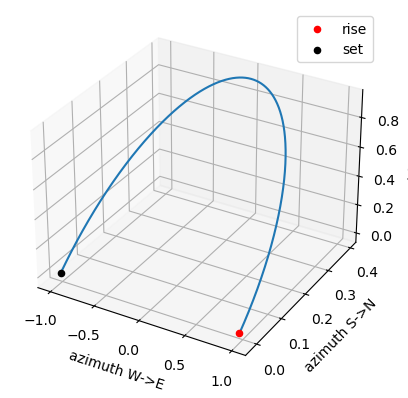}\includegraphics[width=0.49 \textwidth]{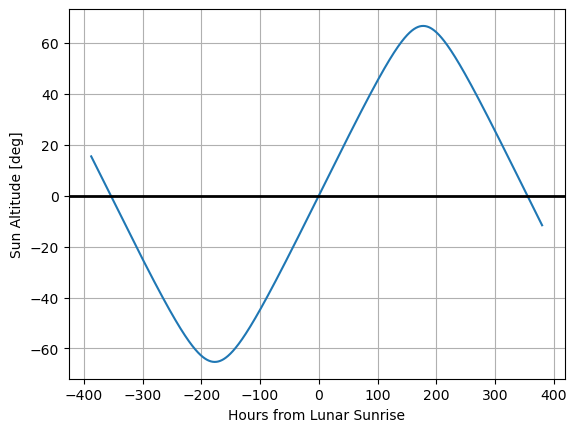}
\end{center}
\caption[]{(Left) Illustration of the trajectory of the sun across the lunar sky at the LuSEE-Night landing site in arbitrary units. The sun is in the northern half of the sky at all times, so full illumination is provided from the north, and shadows are exclusively cast to the south. (Right) Solar altitude as a function of hours from sunrise.}
\label{fig:solar_track} 
\end{figure}

The path of the sun was calculated in altitude and azimuth for the LuSEE-Night landing site at Lunar latitude $-23.815^\circ$, longitude $182.25^\circ$, for the first lunar cycle of 2026, when LuSEE-Night is scheduled to land on the lunar far-side. The obliquity of the moon with respect to the ecliptic plane is only 1.54 degrees, so this solar track will vary only slightly as a function of the time of year, and this variation will not have a significant effect on the calculated power. The solar track for this lunar cycle is shown in Figure \ref{fig:solar_track}, along with the solar altitude as a function of time.

LuSEE-Night is modeled as an enclosure with PV panels on the Top, East, and West faces. The solar power incident on the PV panels is calculated in discrete timesteps of 15~minutes, using a solar constant of $1361 \ \textrm{W}/\textrm{m}^2$, and taking into account the projected area of the PV panel for the solar position at that time step. Additionally, during sunrise and sunset we take into account the reduction in solar power due to the fraction of the solar disk that is below the horizon. This effect is noticeable because sunrise and sunset take \aprx1\,hr on the lunar surface, and the battery charging profile and communications system up-time are both sensitively dependent on the power profile around dawn and dusk.

Further factors are used to calculate the power generated by the PV panel:

\begin{itemize}
\item The active area of the PV panel
\item The nominal room temperature photovoltaic efficiency of the panel
\item The thermal coefficient of the PV efficiency.
\item A derating factor for dust obscuration
\item A derating factor for radiation degradation
\item For the PV panels on the top face: a time dependent factor for reduction of the illuminated PV area due to shadowing by the LuSEE-Night antenna structure. This factor is also dependent on the layout of the top face PV panels.
\end{itemize}

\begin{figure}[H]
\begin{center}
\includegraphics[width=0.75 \textwidth]{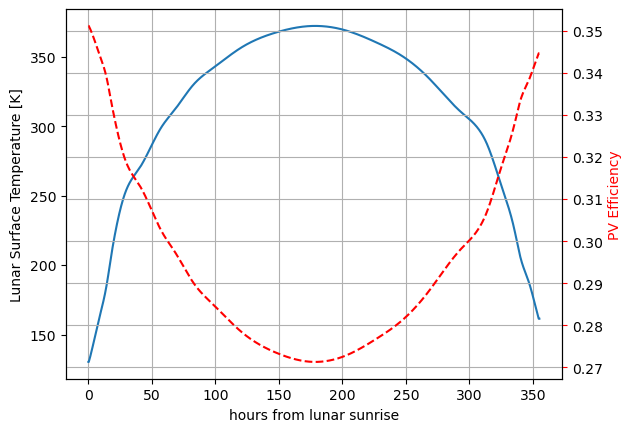}
\end{center}
\caption[]{The solid blue curve shows the temperature at the lunar surface at approximately the LuSEE-Night observatory latitude, as a function of hours from Lunar dawn, as measured by the LRO Diviner Lunar Radiometer Experiment. We have interpolated the measured data to 15\,min. time intervals during the lunar day, to match the timestep of our simulations. The dashed red curve shows the resulting PV thermal efficiency factor, using the specifications given by the vendor.}
\label{fig:lunar_temp} 
\end{figure}

The solar panel vendor provided measurements of the PV panel efficiency at a range of temperatures, as well as at the beginning of life (BOL) and (EOL). These measured parameters were used to create a model of the PV efficiency at the range of operational temperatures that will be experienced on the moon, both at BOL, and derated at EOL.
We then use lunar surface temperature data, as measured by the LRO Diviner Lunar Radiometer Experiment, and interpolated down to 15\,min. increments, to calculate our thermal efficiency as a function of time throughout the lunar cycle. Figure \ref{fig:lunar_temp} shows the measured and interpolated temperatures, and the resulting PV efficiencies, as a function of time. 

For dust obscuration and radiation degradation we assume conservative factors of $5\%$ each. We do not anticipate there will be significant dust obscuration because the main source of dust for planetary landers is dust that is lofted by the landing rockets, and in the vacuum environment of the lunar surface, this dust will follow parabolic trajectories away from the lander. Furthermore, there is no wind to deposit dust on the lander at later times, as there is for landers on Mars. Electrodynamic effects may loft lunar dust, particularly at dawn and dusk, but it is not expected that significant amounts of dust will be deposited by these effects. Furthermore, this effect will be mitigated by a thin conductive layer on the surface of the PV panels, which is designed to prevent electrostatic charge buildup, without significantly affecting the efficiency of the PV cell. Under these conditions, $5\%$ of the PV area being obscured by dust at the End of Life (EOL) for the experiment is considered to be a conservative estimate. Similarly, $5\%$ degradation in PV efficiency due to radiation at EOL is a conservative estimate given the radiation environment on the lunar surface.

\subsection{Top Panel Layout}

\begin{figure}[H]
\begin{center}
\includegraphics[width=0.49 \textwidth]{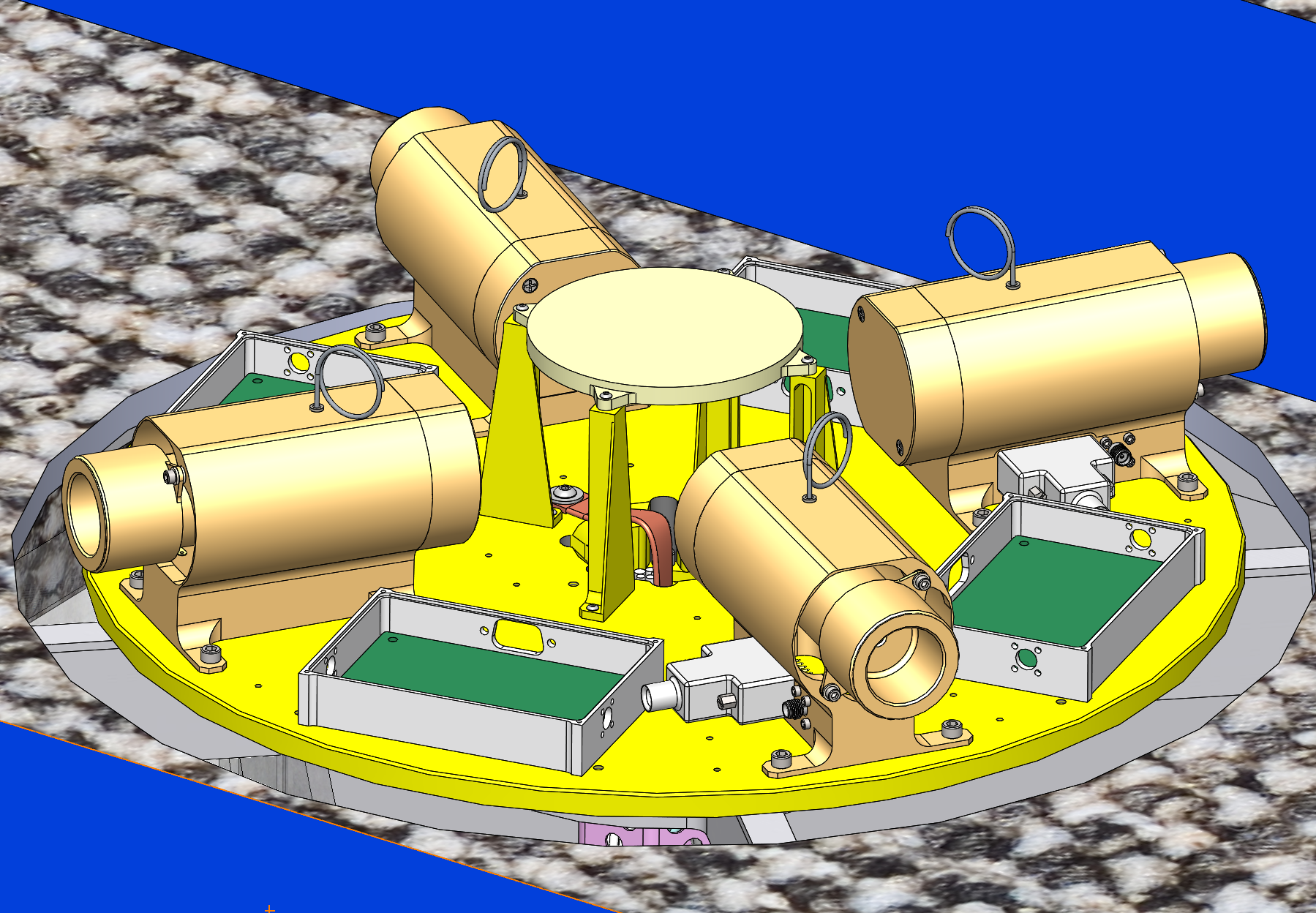}
\end{center}
\caption[]{The LuSEE-Night turntable, with four science antenna deployers (gold color), and the S-band antenna (center). The antenna deployers and the S-band antenna are the largest structures above the PV panels on the top face of the LuSEE-Night instrument, and therefore are responsible for most of the shadow area on the PV panels. The science antennas themselves are significantly smaller diameter than the deployers, and their shadow contributions will be subdominant. The preamplifier enclosures (silver and green) are low profile, and their shadows are entirely contained within those of the science and S-band antennas.}
\label{fig:lusee_ant} 
\end{figure}

There are several structures on the top face of the LuSEE-Night enclosure that can cast shadows over the PV panels on the top face, reducing their illuminated area. These include the four monopole science antennas, the deployer units for the science antennas, the turntable for azimuthal rotation of the monopole antennas, and the S-band antenna for communications and data uplink to the communications relay satellite. The turntable, S-band antennas, and science antenna deployers can be seen in Figure \ref{fig:lusee_ant}. The geometry of these structures is fairly complicated, leading to complicated shadow geometries on the top face PV panels. However, these structures can be simplified in a manner that yields conservative estimates of shadow areas.

First, the deployer modules and S-band antenna are significantly larger than the science antennas, so the shadows cast by the science antennas will be subdominant to those cast by the deployers and S-band antenna. Therefore we neglect the science antennas in our shadowing simulations. Secondly, the exact shape of the deployers and antenna do not significanly impact the shape of their shadows. To simplify the simulation, we therefore model the four antenna deployers and S-band antenna as two crossed plates, oriented vertically, with height equal to the deployer height (104~mm), and length equal to the end-to-end length of the deployer modules, across the turntable (400~mm). We calculate the shadow dimensions cast by these two plates at each timestep in our simulation. Then, given a PV geometry, discretized into 1~cm $\times$ 1~cm sections, we calculate which PV pixels are shadowed at each timestep, and the total shadowed fraction. The crossed plates are nominally oriented N-S and E-W. During the morning and evening the N-S plate will be the primary source of shadows, to the west and east respectively, while during the midday the E-W plate will be the primary source of shadows, which will be cast exclusively on the southern half of the top face of LuSEE-Night. However, due to the high altitude of the sun around midday, the shadows to the south are quite short, while the shadows at dawn and dusk cover most of the west and east portions of the top face. The preamplifier enclosures are significantly lower profile than the deployers and S-band antenna, and their shadows are therefore almost entirely enclosed within those of the other components, and are neglected in our simulations for simplicity.

For these reasons, a PV layout with two bands along the north and south edges of the top face, where little or no shadowing occurs, would be optimal for PV efficiency. However, the PV layout can also affect the instrument beams, since the conductivity of the PV panels is different from that of the metal top face, which acts as a back-plane to the science antennas, and helps to shape their beams. We performed electromagnetic simulations of a two-stripe PV design using the commercial HFSS and FEKO software, and we find that the asymmetric nature of this layout negatively impacts the science beam symmetry. Therefore, an azimuthally symmetric PV layout is preferable for the LuSEE-Night science. 

\begin{figure}[H]
\begin{center}
\includegraphics[width=0.80\textwidth]{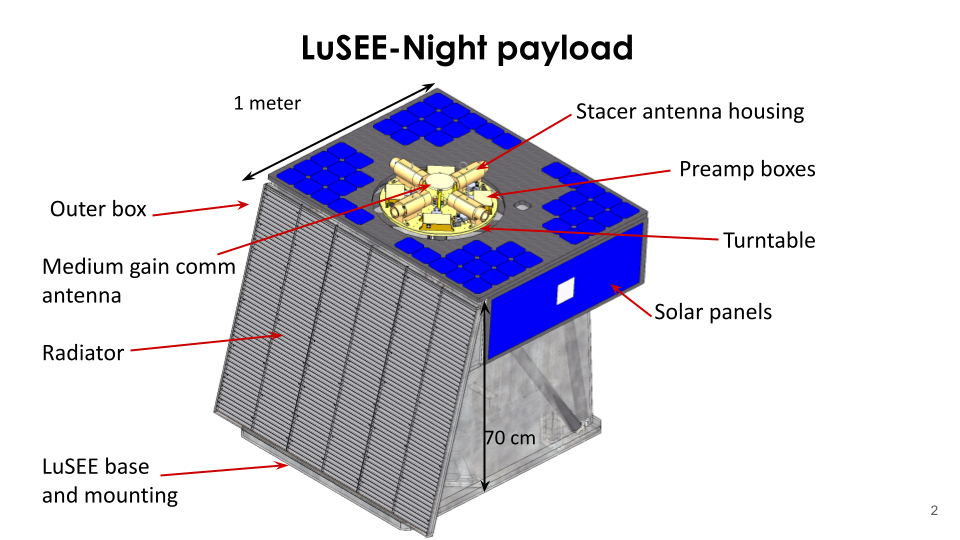}
\end{center}
\caption[]{The final PV layout for LuSEE-Night consists of four strings of 13 Coverglass Interconnected Cells (CICs) arranged around the outer border of the top face, with an open margin between the turntable and the CICs. The antenna turntable is centered on the lander and base of the LuSEE-Night enclosure for mechanical stability and beam symmetry, but slightly offset with respect to the PV panels due to limited area atop the enclosure. The approximate azimuthal symmetry of this layout simultaneously minimizes the effects of the PV system on the science antenna beam pattern, and minimizes the power generation loss due to shadowing.}
\label{fig:pv_layout} 
\end{figure}

Our final PV layout then simultaneously minimizes effects on the science beam, and the area of shadows on the PV. This layout consists of PV cells arranged symmetrically around the entire top face, with an empty annular margin immediately around the turntable, where there is greater shadowing. This layout is shown in Figure \ref{fig:pv_layout}. Due to limitations in the area on the top of the instrument enclosure, and the need to keep the antenna rotation stage centered on the lunar lander for mechanical stability and beam symmetry, the PV pattern is slightly offset with respect to the antennas. This asymmetry is substantially smaller than our minimum wavelength, and does not produce a significant effect on the science beam pattern.

\subsection{Side Panel Layout}

Locating the PV panels on the top face of the LuSEE-Night enclosure maximizes the power generated per unit area, averaging over the day cycle. The disadvantage of locating the PVs exclusively on the top face however, is that the projected area is very low at dawn and dusk. Given the limited mass budget for battery in LuSEE-Night, and the long (328\,hr) duration of the lunar night, the battery will be significantly discharged by dawn. Furthermore, we have a limited data rate at which we can transmit our science and housekeeping data through the relay satellite back to Earth. We would therefore like to maximize the amount of time that our communications system is operational during the lunar day, to maximize the amount of data returned for analysis and operations. In order to begin recharging the battery as soon as possible after dawn, we consider adding PV panels to the east face of the LuSEE-Night enclosure. Likewise, in order to continue charging the battery as long as possible at dusk, to keep the comms system active and keep the battery full to maximise nighttime observing, we consider adding a PV panel to the west face of the enclosure.

\begin{figure}[H]
\begin{center}
\includegraphics[width=0.49 \textwidth]{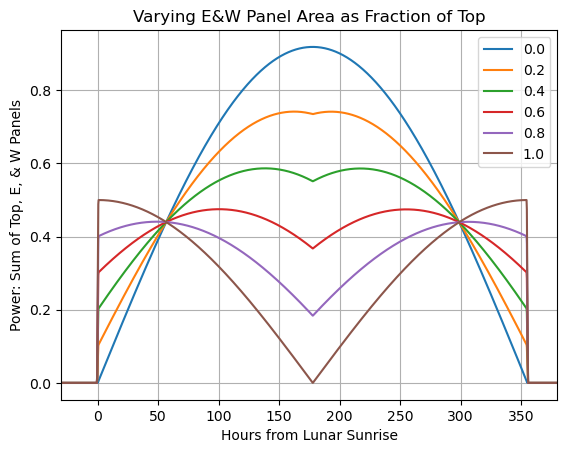}
\includegraphics[width=0.49 \textwidth]{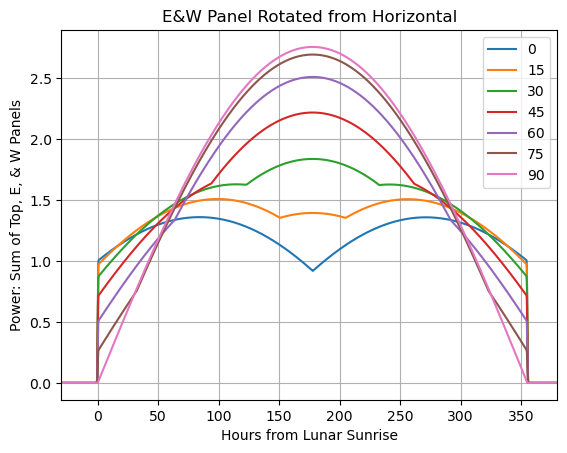}
\end{center}
\caption[]{(Left) The effects of varying the fraction of total panel area on the sides of the instrument enclosure vs the top face. Between 0.4 and 0.6 gives a good balance between total power generated and crucial dawn/dusk power. (Right) The effects of varying the angle of the side panels, for normal vectors from $0^\circ$ with respect to horizontal, all the way to $90^\circ$. This is essentially degenerate with varying the side fraction, and we set the normal vector as $0^\circ$ for mechanical simplicity.}
\label{fig:EW_panel_area} 
\end{figure}

There are two tradeoffs to be simulated in the design of the side panels. First, for a fixed total area of PV panels, we simulated varying the fraction of panel area on the side faces vs the top face. Figure \ref{fig:EW_panel_area} shows the results. The side panels will be less efficient averaged over the lunar day, but provide crucial power when the sun is at low altitudes. A layout with $50\%$ of the total panel area distributed on the side panels was found to provide a balance between total power generated and dawn/dusk power. Second, one could rotate the normal vector of the side panels from horizontal towards vertical, to balance between total power generated and dawn/dusk power (See also Figure \ref{fig:EW_panel_area}). This is essentially degenerate with varying the side panel area fraction, and we selected a normal vector oriented at $0^\circ$ with respect to the horizon (vertical PV panels) for mechanical simplicity.

\subsection{Lunar Lander Alignment}

\begin{figure}[H]
\begin{center}
\includegraphics[width=0.49 \textwidth]{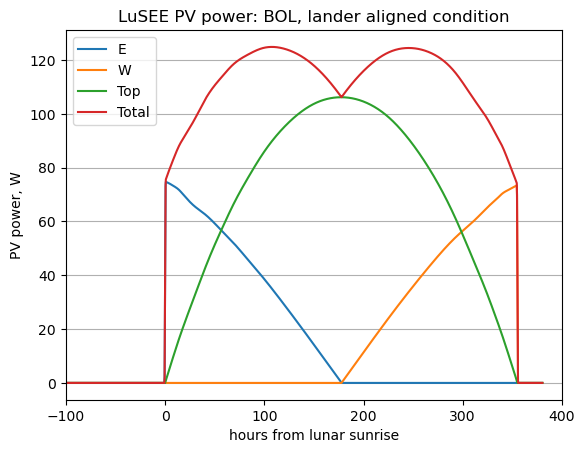}
\includegraphics[width=0.49 \textwidth]{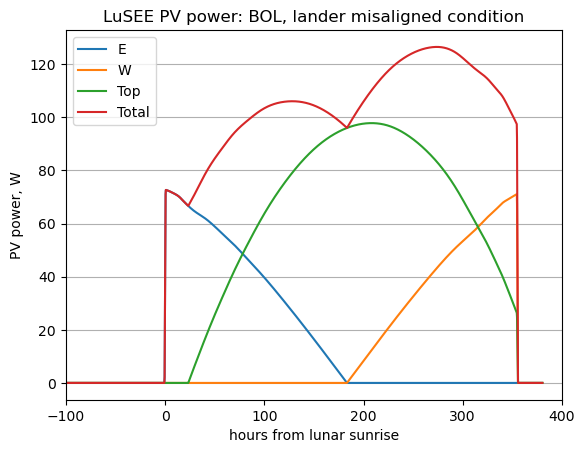}
\end{center}
\caption[]{(Left) Power received by the East, West, and Top PV arrays, and total power, with the lander correctly aligned with the cardinal directions on landing. (Right) PV power generated with the lander maximally misaligned, by $10^\circ$ in each axis. Here the lander is oriented such that all panels are less efficient (normal vector is farther from the sun at optimal illumination time), and the time of peak power generation is shifted towards nightfall. While the decrement in total power generated ($10\%$) is small enough that the battery can still be fully charged, the difference in power profile will affect when daytime operations can optimally be scheduled.}
\label{fig:misaligned_lander} 
\end{figure}

In the preceding analyses we assume that the lander is perfectly aligned with the cardinal directions, which is the intended landing orientation. It is possible, however, that the lander will not be perfectly aligned upon landing. Firefly Aerospace has given us specifications stating that the lander will be aligned with the cardinal directions to within $\pm10^\circ$ in all axes, and that after landing they will verify the lander orientation, and report it to the LuSEE-Night science collaboration to within $0.1^\circ$. Knowing the misalignment of the lander will allow us to correct for our science beam pointing, but it will also affect the power generation by the PV arrays. We therefore calculated the power generated as a function of lander rotation in three axis, and found the worst case decrement in power generated, at the maximum allowed misalignment of $10^\circ$ in each axis, to be $10\%$. In our analysis this is less than the excess energy provided by the PV array over the energy needed to charge the battery during the day. Therefore our selected PV array layout has mitigated the risk of lander misalignment for power generation, up to our specification of $10^\circ$ maximum misalignment in each axis. Knowing the actual lander alignment will still be important for informing the charging profile during the lunar day, and scheduling the timing of operations during the lunar day (See Figure \ref{fig:misaligned_lander}.)

\section{Power Management and Concept of Operations}

For simulating overall power management, including power accumulated throughout a lunar day, and available instrument power at night, we can use a simplified instrument power model which assumes all components have a constant power draw when active. This allows us to estimate our total nightly instrument runtime, and daily data transfer amount. For planning the timeline of operations during the lunar day and night, our concept of operations or ``ConOps", we use a more granular approach in which we track the variable power loads of instrument components over time. These models will both be discussed below.

\subsection{Power Management with Constant Loads}

In this section we present a simplified instrument power balance during the lunar cycle. 
For power sources, we assume the battery is nearly fully charged to its nameplate capacity at the beginning of each lunar night. The battery's state of charge (SOC) runs down during the night at a constant rate. 
During the day, when in a power-positive condition, spectrometer observations will re-start and the S-band radio will make periodic contact with the relay satellite. Solar array power in excess of what is needed to operate the spectrometer and radio will recharge the battery. 

In this analysis, the solar array power generation is modeled assuming no lander misalignment, but includes the effect of temperature on panel efficiency as shown in Figure \ref{fig:lunar_temp}, and the derating factors from Section \ref{sec:solar_sims}. The electrical loads on the system are modeled as constant power loads, one value each for night and day operations. This model is expanded Section \ref{sec:conops}, to include time variations in the various electrical loads. The power loads included in the calculation are:

\begin{itemize}
    \item Peak Power Tracker (PPT): 1.8~W
    \item Power Distribution Unit (PDU):  0.1~W
    \item Picket Fence Power Supply (PFPS): 6~W 
    \item Digital Control Board (DCB): 2~W
    \item Spectrometer Module (SPT): 9.25~W 
    \item Preamplifier Modules: 0.86~W
    \item Radio Frequency Receiver (RF RX): 5.3~W
    \item Radio Frequency Transmitter (RF TX): 13.7~W
\end{itemize}

\begin{figure}[H]
\begin{center}
\includegraphics[width=0.75 \textwidth]{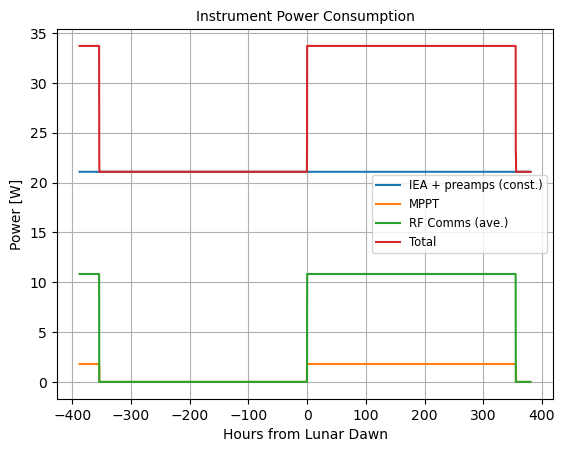}
\end{center}
\caption[]{The LuSEE-Night expected instrument power consumption. The spectrometer instrument, preamplifiers, and associated power systems will operate throughout the lunar night. During the day, the peak power tracker will monitor and operate the solar array, and the RF communications system will relay science and housekeeping data to the communications relay satellite.}
\label{fig:power_consumption} 
\end{figure}

We also include charging and discharging efficiencies of $95\%$, and an uncertainty margin on the total power load of $10\%$. Figure \ref{fig:power_consumption} shows the total instrument power consumption including these factors. During the lunar night only the core instrument will be operational (the PDU,  PFPS, DCB, SPT, and Preamplifiers). During the day, the PPT will be active to monitor and operate the solar array, and the RF RX and TX modules will periodically be active to send and receive data through the relay satellite. Here the RX and TX modules are modeled as always on, as a conservative estimate of power requirements. This will be refined to only have the TX module on during relay satellite passes in Section \ref{sec:conops}.

\begin{figure}[H]
\begin{center}
\includegraphics[width=0.75 \textwidth]{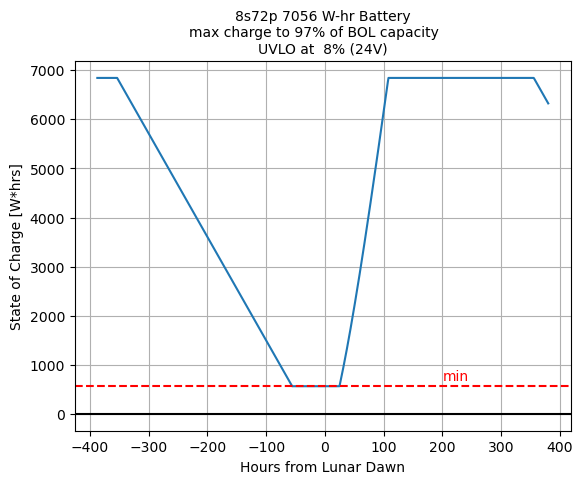}
\end{center}
\caption[]{The LuSEE-Night battery is expected to begin each lunar night fully charged to the nameplate capacity, and will slowly discharge throughout the 328 hours of night as it powers the spectrometer and other night operations systems. Here we show that it will be necessary to duty cycle the spectrometer, as there is insufficient stored power for continuous observations. Daytime operations will commence as soon as the solar array has sufficient power to run all onboard systems, and battery charging will begin when there is excess power available.}
\label{fig:soc} 
\end{figure}

The resulting expected state of charge (SOC) of the LuSEE-Night battery is shown in Figure \ref{fig:soc}. The minimum safe battery bus voltage, at which the battery can be safely recharged, is taken here to be 24~V, which occurs at $8\%$ SOC. Due to the available battery capacity (7~kW-hrs), limited by the mission mass specifications, we find that LuSEE-Night will not be able to operate continuously throughout the entire lunar night. 
To prevent the battery from discharging below a safe bus voltage it will be necessary to duty-cycle the spectrometer during the night to preserve power. This is simulated in the time-variable power simulations. Duty cycling the spectrometer may also be scientifically advantageous, so that data can be taken at uniformly distributed times throughout the night.

\begin{figure}[H]
\begin{center}
\includegraphics[width=0.75 \textwidth]{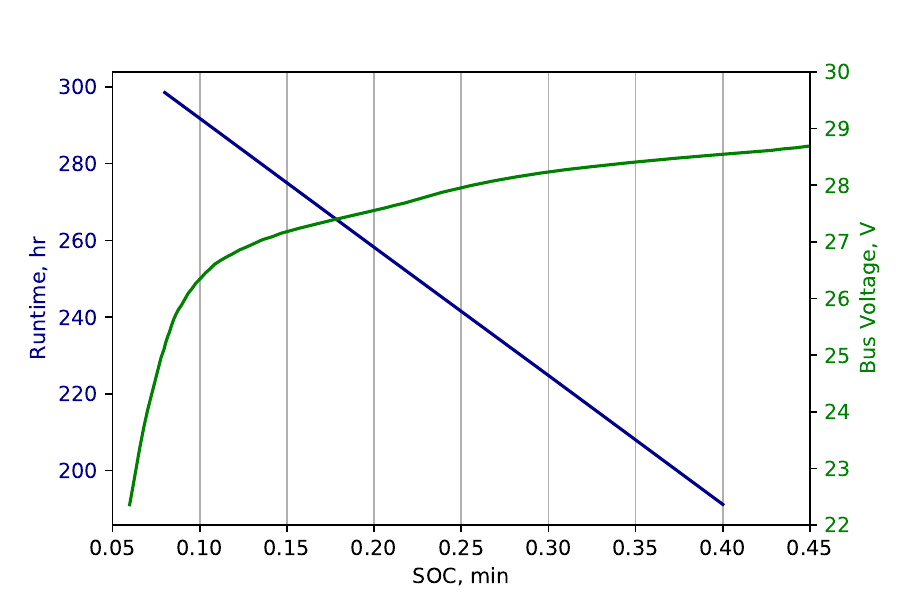}
\end{center}
\caption[]{Blue: battery run time to support full nighttime operation vs. minimum SOC. Green: battery bus voltage vs. SOC.}
\label{fig:runtime} 
\end{figure}

The 24~V minimum voltage ($8\% SOC$) shown in Figure \ref{fig:soc} is also an aggressive figure. Limiting the battery to a more conservative minimum SOC will further reduce the available runtime. Figure \ref{fig:runtime} illustrates the trade off. For the ConOps plan developed for LuSEE-Night, we set our minimum SOC to be $30\%$, for a minimum bus voltage of \aprx28~V.

During the day, the solar array has sufficient power generation capacity to fully recharge the battery to its nameplate capacity, while allowing for RF RX and TX operations throughout the day, and spectrometer operations while the RF TX/RX systems are not active, in between relay satellite passes. 
We expect that the battery can be fully recharged within $\aprx 100$ hours of charging time, leaving time for daytime observations, and a large safety margin to compensate for potential decreased charging efficiency and degraded power generation as the operational life of the battery and solar array are extended.

\subsection{Concept of Operations with Variable Power}
\label{sec:conops}

The LuSEE-Night instrument is controlled by the onboard flight computer, or Data Control Board (DCB). The DCB monitors the state of the instrument and controls the onboard power systems, communications with the relay satellite, data storage and transfer, and interfaces with the spectrometer. The DCB can receive instructions during the day, but must control the instrument autonomously during the night. Therefore, it is necessary to have a rigorously tested concept of operations, by which the DCB will operate the instrument. To facilitate the ConOps, we define four operational modes which encompass all normal operations for LuSEE-Night. The task of ConOps is then reduced to switching the instrument between the different modes of operation. Operations over multiple lunar cycles can be simulated, and a ConOps table developed which will determine the appropriate instrument operational mode depending only on the time in the lunar cycle. These simulations take into account the time in the cycle, the SOC of the battery, the instrument internal temperature, data storage volume, and (during the day) the current power being generated by the PV array and whether the relay satellite is expected to be within view. In addition to the normal modes of operation, there are also two ``safe-modes'' which the DCB can enter if the monitored conditions diverge from those in the planned sequence of operations. These safe-modes prioritize survival of the instrument, and re-establishing communications during the next lunar day, so that the instrument can be debugged, and an updated ConOps table transmitted to the DCB.

\begin{figure}[H]
\begin{center}
\includegraphics[width=1.0 \textwidth]{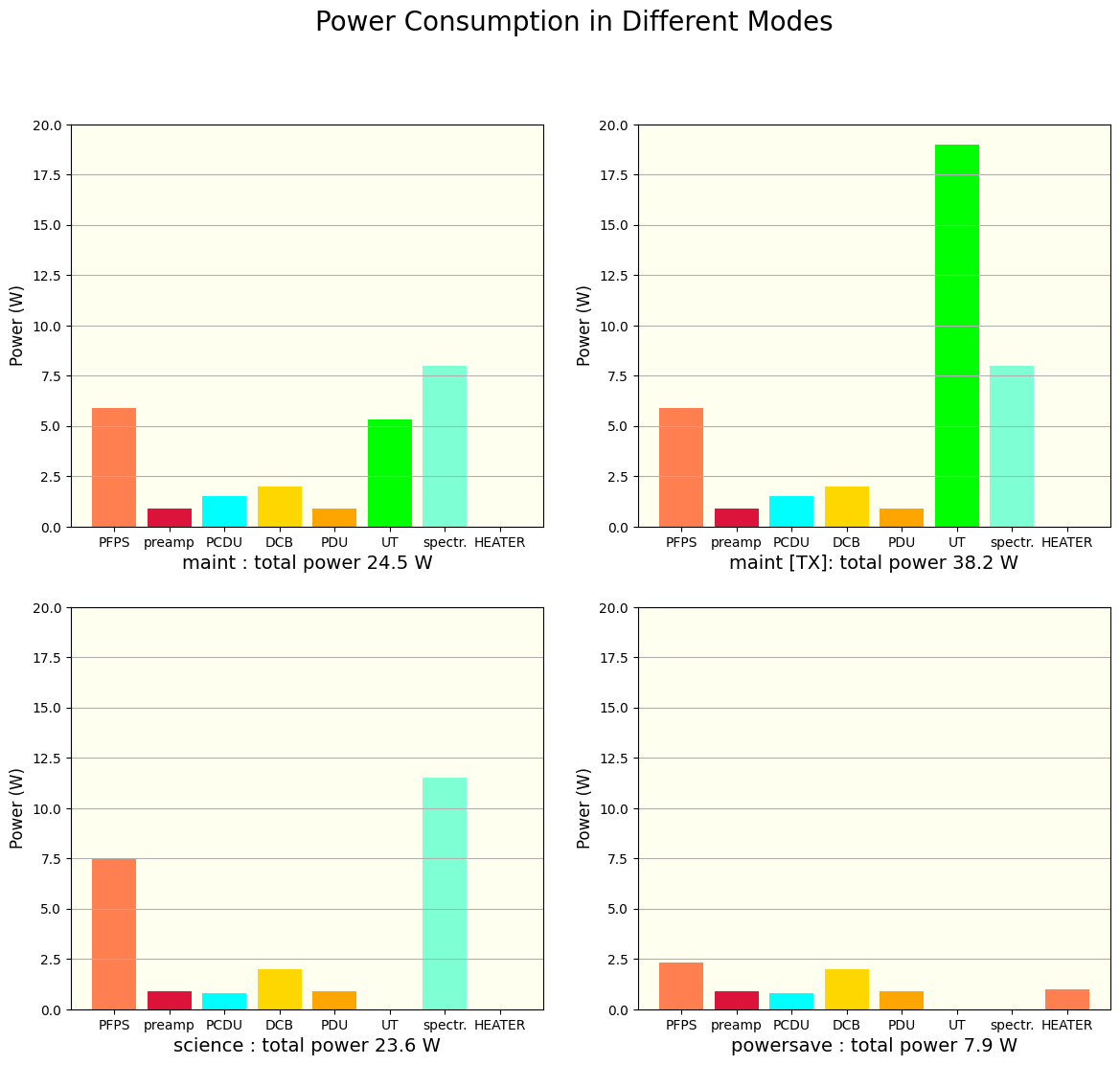} 
\end{center}
\caption[]{Component power consumption for different modes of operation for LuSEE-Night. From top-left to bottom-right:  a) Maintenance mode, the default mode during daytime, with communications in receive mode b) Maintenance [Transmit] mode, when the S-band communications antenna is transmitting data to the relay satellite during the day c) Science mode, the default operational mode during nighttime observations, and d) Powersave mode, with the spectrometer disabled to conserve battery power during the night, and a low power heater enabled to keep the battery at operational temperature.}
\label{fig:modes} 
\end{figure}

The four operational modes and their associated power draws are shown in Figure \ref{fig:modes}. The DCB and power supply systems are necessarily active in all modes of operation. Additionally, the science antenna pre-amplifiers are always active because they are more stable if left powered at all times, and their power draw is minimal. The power states of the other components vary between modes as follows:

\begin{itemize}
    \item Maintenance Mode - The default mode during the daytime. The communications module (UT) is powered on in receive mode, waiting for a ping from the communications relay satellite, indicating that it is within range. The spectrometer is active for daytime observations and calibration.
    \item Transmission Mode - The mode during active communication with the relay satellite. The transmitter (TX) is powered on, increasing power draw and waste heat.
    \item Science Mode - One of two nighttime modes. The UT is powered off to prevent self-generated RFI, and the spectrometer is powered on for data taking.
    \item Powersave Mode - The spectrometer, and additional power to the PFPS to run it, are the main power draw in Science Mode. To conserve power during the night, the spectrometer may be powered off. A small heating element is enabled to keep the battery temperature within operational range.
\end{itemize}

\begin{figure}[H]
\begin{center}
\includegraphics[width=1.0 \textwidth]{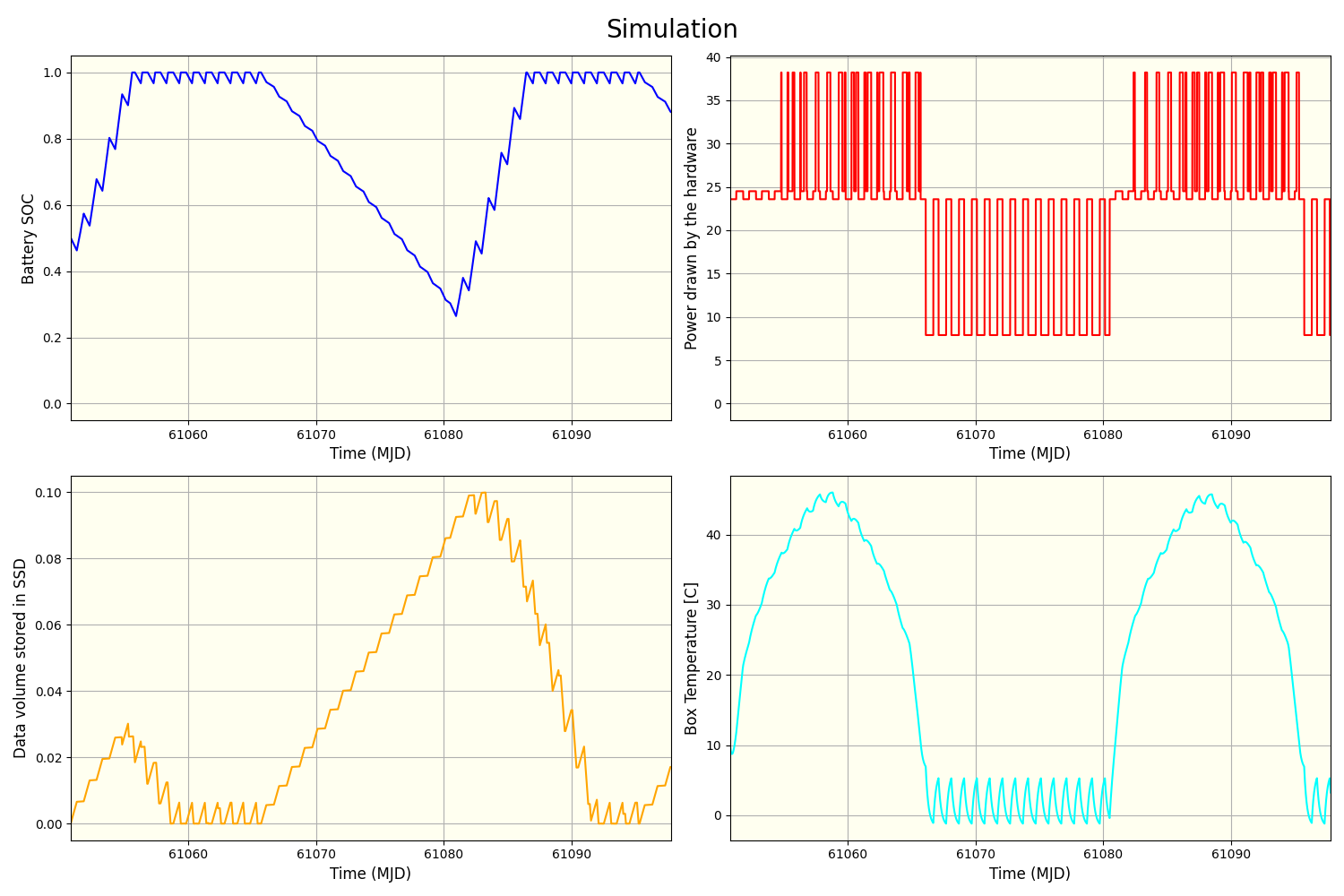} 
\end{center}
\caption[]{From top-left to bottom-right, simulated a) battery SOC, b) instrument total power draw, c) data storage volume, and d) enclosure internal temperature, using variable power rates for all components, given by the operational modes displayed in Fig. \ref{fig:modes}, and using the LuSEE-Night ConOps software to manage the switching between operational modes to simultaneously manage SOC, data volume, and temperature.}
\label{fig:conops} 
\end{figure}

Figure \ref{fig:conops} shows the results of simulations using these power modes and the planned ConOps table, and tracking battery SOC, power draw, data storage volume, and enclosure internal temperature as a function of time, during the day and night cycle.

During the night, battery SOC is depleted, and data is accumulated in the onboard SSD. Power draw oscillates as the spectrometer is duty cycled to ensure adequate battery power is preserved until dawn. During the day, SOC increases from solar power charging the battery, and data volume stored decreases as data is transmitted to Earth with each relay satellite pass. The spectrometer is always active, but power draw oscillates as the relay satellite passes overhead and the TX unit is active. Note that the cadence of these power oscillations are predictable, but irregular, due to the irregular orbital pattern of the LPF relay. Note that TX operations do not commence immediately at dawn, but wait for a safe SOC to be reached first. 

The battery charging is paused intermittently during the day, while day-time science observations are being taken. The PV system's Maximum Point Power Tracker (MPPT) is not powered from the PFPS, and therefore has to be disabled while observing to prevent self-generated RFI. The battery is still expected to reach maximum SOC in less than half of the lunar day. 

Data volume increases during the day in between relay passes because the spectrometer is still active. Data gathered during the day is noisier, due to radio frequency radiation from the sun, and therefore less useful for cosmological analysis. However, it is still expected to be useful for secondary science (such as solar or galactic physics) and for calibration purposes. 

Internal temperature increases during the day, especially during periods when the TX unit is active, but is managed by radiating heat away to vacuum with a custom parabolic radiator panel. At night a thermal switch connecting to the radiator is opened to cut off the connection to the radiator, and a safe operating temperature is maintained by a combination of insulation and self-generated heat from the instrument. At night, temperature increases while the spectrometer is active, and decreases when it is inactive. However, cooling in powersave mode is slowed with a low powered heating element, as well as by waste heat from the other electronics components that must remain active at all times, including the DCB and power supply systems.

During the first lunar night, a more conservative operations schedule will be used, with less observing time and more time in powersave mode, to ensure that a safe margin of battery power remains at lunar dawn. As an additional safety measure, there will will also be three nighttime communications intervals to check in on the health of the instrument, and verify that it is performing as expected. These will be i) Several hours into the night, after thermal equilibrium is expected to be reached ii) One day into the night, after a significant amount of science data has been recorded, and iii) Half way through the lunar night. These communications sessions will be focused on retrieving crucial housekeeping data first, and perhaps a small amount of science data afterwards. Thereafter, once the electronic systems and ConOps software are tested in situ, the standard operations schedule will be employed for subsequent nights, and all communications will be restricted to lunar days.

\section{Summary}

This paper has described the design of the LuSEE-Night power system hardware, including the PV array geometry and layout, and the design of the power management and ConOps software. The design and operation of these systems are essential for the success of the LuSEE-Night experiment, which is attempting an unprecedented multi-night observation campaign in the harsh environment of the lunar far-side. For these observations to succeed, the instrument must be powered throughout the night and recharged during the day, it must maintain thermal equilibrium in its operational range, and it must transmit the instrument's scientific and housekeeping data back to Earth for analysis. The power systems, and the ConOps software which will control the power systems, spectrometer, and communications have been thoroughly simulated. Electrical, thermal, and vacuum testing will also be performed to verify the simulations in the laboratory environment before the instrument is deployed. And once on the moon, a period of testing and more frequent communications is planned at the opening of observations to verify performance in situ. Together with the other instrument systems, these simulations and testing of the power and control systems will allow LuSEE-Night to open a new observational window into the Cosmic Dark Ages, and break ground on a revolutionary new low-frequency radio observatory location, on the far side of the moon.

\acknowledgments 
 
This work was supported by the U.S. DOE Office of High Energy Physics, under Contract No. DE-AC-0205CH11231, and by NASA under contract  No. 80MSFC23CA015 to the University of California, Berkeley. 

\bibliography{lusee_bib} 

\begin{thebibliography}{10}

\bibitem{Chang:2007xk}
Chang, T.-C., Pen, U.-L., Peterson, J.~B., and McDonald, P., ``{Baryon Acoustic Oscillation Intensity Mapping as a Test of Dark Energy},'' {\em Phys. Rev. Lett.}~{\bf 100},  091303 (2008).

\bibitem{deboer17}
{DeBoer, David R. et al.}, ``{Hydrogen Epoch of Reionization Array (HERA)},'' {\em Publications of the Astronomical Society of the Pacific}~{\bf 129},  045001 (Mar 2017).

\bibitem{chime22}
Amiri, M. and the Chime~Collaboration, ``An overview of chime, the canadian hydrogen intensity mapping experiment,'' {\em The Astrophysical Journal Supplement Series}~{\bf 261},  29 (July 2022).

\bibitem{crichton22}
{Crichton, D et al.}, ``Hydrogen intensity and real-time analysis experiment: 256-element array status and overview,'' {\em Journal of Astronomical Telescopes, Instruments, and Systems}~{\bf 8} (Jan. 2022).

\bibitem{oconnor20}
{O'Connor, P. et al.}, ``{The Baryon Mapping Experiment (BMX), a 21cm intensity mapping pathfinder},'' {\em Society of Photo-Optical Instrumentation Engineers (SPIE) Conference Series} {\bf 11445} (Dec. 2020).

\bibitem{bandura19}
{Bandura}, K. and {Castorina}, E.~{Connor}, L., ``Packed ultra-wideband mapping array (puma): A radio telescope for cosmology and transients,'' {\em Bulletin of the American Astronomical Society}~{\bf 51} (2019).

\bibitem{Monsalve19}
Monsalve, R.~A., Fialkov, A., Bowman, J.~D., Rogers, A. E.~E., Mozdzen, T.~J., Cohen, A., Barkana, R., and Mahesh, N., ``Results from edges high-band. iii. new constraints on parameters of the early universe,'' {\em The Astrophysical Journal}~{\bf 875},  67 (Apr. 2019).

\bibitem{philip19}
Philip, L., Abdurashidova, Z., Chiang, H.~C., Ghazi, N., Gumba, A., Heilgendorff, H.~M., Hickish, J., Jáuregui-García, J.~M., Malepe, K., Nunhokee, C.~D., Peterson, J., Sievers, J.~L., Simes, V., and Spann, R., ``Probing radio intensity at high-z from marion: 2017 instrument,'' {\em Journal of Astronomical Instrumentation}~{\bf 08} (2019).

\bibitem{singh21}
Singh, S., T., J.~N., Subrahmanyan, R., Shankar, N.~U., Girish, B.~S., Raghunathan, A., Somashekar, R., Srivani, K.~S., and Rao, M.~S., ``On the detection of a cosmic dawn signal in the radio background,'' {\em Nature Astronomy}~{\bf 6} (2021).

\bibitem{tamura24}
{Tamura}, E., {Fried}, J., {Herrmann}, S., {O'Connor}, P., {Raguzin}, E.~J., {Slosar}, A., and {Bale}, S.~D., ``{Development and Characterization of the Flight Model Spectrometer Onboard LuSEE-Night},'' {\em Society of Photo-Optical Instrumentation Engineers (SPIE) Conference Series} (June 2024).

\bibitem{bale23}
Bale, S.~D., Bassett, N., Burns, J.~O., Jones, J.~D., Goetz, K., Hellum-Bye, C., Hermann, S., Hibbard, J., Maksimovic, M., McLean, R., Monsalve, R., O'Connor, P., Parsons, A., Pulupa, M., Pund, R., Rapetti, D., Rotermund, K.~M., Saliwanchik, B., Slosar, A., Sundkvist, D., and Suzuki, A., ``Lusee `night': The lunar surface electromagnetics experiment,'' {\em URSI GASS} (2023).

\end{thebibliography}
\bibliographystyle{spiebib} 

\end{document}